\documentclass[aps,prd,twocolumn,superscriptaddress,nofootinbib,longbibliography]{revtex4-2}
\usepackage{amsmath,amssymb}
\usepackage{booktabs}
\usepackage{graphicx}
\usepackage{microtype}
\usepackage{float}
\graphicspath{{../figs/}{figs/}{./}}

\begin{document}

\title{Longitudinal-dither wavefront sensing:\\
Calibration-free measurement of mode mismatch}
\author{Masayuki Nakano}
\email{masayuki.nakano@ligo.org}
\thanks{ORCID 0000-0001-7703-0169}
\affiliation{LIGO Laboratory, California Institute of Technology,
Pasadena, CA 91125, USA}
\date{\today}

\begin{abstract}
Mode mismatch between a gravitational-wave detector and the beams injected
into it lowers the power stored in the arms and the level of squeezing.
Correcting it requires the complex amplitudes of the three second-order
modes, measured on a calibrated scale with the cavity on resonance at its
operating power. Here we present a method that obtains a signal proportional
to the mismatch by moving the beam on the wavefront sensors already used for
alignment control. A longitudinal dither of the cavity adds to the resonant
mode a reference field, which reaches the sensors through the same signal
chains and so calibrates both the scale of this signal and the Gouy phases of
the sensors and actuators, with no external calibration. We demonstrate the method on an optical cavity held on
resonance in its operating state at input powers of $2$\,W and $62$\,W. At
$2$\,W it yields an optical model that agrees with independent measurements
in the mode content, and with the layout model in the
actuator Gouy phase within its uncertainty of about $1^\circ$. At $62$\,W the
thermally induced change of the eigenmode is decomposed into waist-size and
waist-position changes. The method requires only
components that every controlled cavity already possesses and enables
continuous in situ monitoring of the mode matching in the operating state.
\end{abstract}

\maketitle

\section{Introduction}
\label{sec:intro}

Current gravitational-wave detectors, such as those of the Laser
Interferometer Gravitational-Wave Observatory (LIGO), are limited by quantum
noise over a wide frequency band~\cite{capote2025}. Shot noise is reduced by raising
the power stored in the arms~\cite{caves1981}, and frequency-dependent
squeezed vacuum now reduces the quantum noise further, radiation-pressure
noise included~\cite{ganapathy2023}. An obstacle to further improvement is the
mode mismatch between the interferometer and the beams injected into it, the
input laser beam and the squeezed vacuum. It lowers the arm power and, for the
squeezed vacuum, acts as an optical loss, resulting in a lower squeezing level.
Mode matching matters in the same way wherever an
experiment relies on storing high power in an optical cavity, as in electron
microscopy~\cite{turnbaugh2021}, neutral-beam injection for
fusion~\cite{fiorucci2022}, extreme-ultraviolet generation~\cite{liu2024},
and axion searches~\cite{heinze2024}.

Written in the eigenmode basis, a mismatched beam contains even-order
Hermite--Gauss modes, and for a small mismatch the second-order modes
dominate~\cite{anderson1984,bayerhelms1984}. The interferometer or the cavity
reflects these modes, so measuring the complex amplitudes $c_{20}$, $c_{02}$,
and $c_{11}$ of the three second-order modes in the reflected light shows what
mismatch the beam has relative to the eigenmode. The amplitude $c_{20}$
describes the mismatch in the horizontal direction, with a waist-size offset in
its real part and a waist-position offset in its imaginary part. The amplitude
$c_{02}$ does the same in the vertical direction, and $c_{11}$ carries the
astigmatism along the diagonal axes. What matters is the mismatch at the
operating power, since the thermal load changes the eigenmode. The measurement
therefore has to run on resonance in the operating state.

In LIGO, mode mismatch is corrected on two scales. Corrective actuators such as thermal
compensation systems (TCS) and piezoelectric suspended active matching stages
(PSAMS)~\cite{brooks2016,srivastava2021} tune the
mode matching in place, but they run without a closed-loop error signal, since
no signal in the present detectors reads these complex amplitudes. Their
range is also limited, so a larger correction needs a change of the optical
layout, moving or exchanging optics inside the vacuum system at the cost of a
vent and the recovery after it. Such a change has to be planned in advance,
from a calibrated measurement that gives the waist-size and the waist-position
offsets separately. Furthermore, a spherical lens or actuator cannot correct
astigmatism, so
knowing the mismatch in each direction, not only its total, tells how far such
a correction can improve the matching. These problems carry over to the
next-generation detectors~\cite{punturo2010,evans2021} and grow more serious
there, as those detectors aim at higher sensitivity.

Several schemes have been proposed to sense the mismatch on resonance, based on
heterodyne detection with a mode converter and radio-frequency (RF) quadrant
photodetectors~\cite{magana2019,schiworski2026}, RF modulation of the input beam shape
read out by a single photodetector~\cite{ciobanu2020,tao2023,tao2025}, phase
cameras~\cite{cabrita2024}, and a sideband injected at a higher-order-mode
resonance and detected with a phase camera~\cite{cabrita2025} or a clipped
photodiode~\cite{goodwinjones2023}. Each of them senses part of the mismatch,
but none has yet measured all six of its degrees of freedom, the real and
imaginary parts of $c_{20}$, $c_{02}$, and $c_{11}$, separately and on a
calibrated scale.

This paper demonstrates such a measurement with existing equipment only, on an
in-vacuum cavity held on resonance at input powers up to $62$\,W, the power at
which LIGO operates~\cite{capote2025}. In this method, moving the beam on the wavefront sensors produces a signal
proportional to the second-order mode amplitude, and a reference field made
by a longitudinal dither of the cavity length calibrates this signal. We call this longitudinal-dither
wavefront sensing (LDWFS).

\section{Method}
\label{sec:method}

\subsection{Configuration and notation}
\label{sec:config}

Consider a single cavity held on resonance with the Pound--Drever--Hall (PDH)
technique~\cite{drever1983}. An electro-optic modulator (EOM) imprints
RF phase-modulation sidebands at the frequency $\Omega$ on the input beam.
At the reflection port, two RF quadrant photodetectors at different Gouy
phases, the wavefront sensors (WFS) A and B, sense the alignment of the
input beam to the cavity~\cite{morrison1994}.

In this configuration, a misalignment or a mismatch of the input beam appears
as higher-order modes in the Hermite--Gauss basis of the cavity eigenmode,
written $u_n$, with $u_0$ the fundamental mode $\mathrm{TEM}_{00}$. We take
both to be small compared with the beam size, the Rayleigh range, and the
divergence angle, so that to lowest order a misalignment adds only the
first-order mode $u_1$, with complex amplitude $a_1$, and a mismatch only the
second-order mode $u_2$, with complex amplitude $c_2$. The phase of each amplitude is taken relative
to $u_0$ at the cavity waist, and since $u_n$ accumulates a Gouy phase $n\eta$
relative to $u_0$, the amplitudes read $a_1 e^{-i\eta}$ and $c_2 e^{-2i\eta}$
at a plane where the accumulated Gouy phase is $\eta$. We treat the horizontal and the vertical plane
separately, each in one transverse dimension, and $c_2$ then stands for
$c_{20}$ in the horizontal plane and for $c_{02}$ in the vertical plane. The
third amplitude, $c_{11}$, which involves both planes, is taken up in
Sec.~\ref{sec:full}.

\subsection{The second-order signal in reflection}
\label{sec:shared}

\begin{figure}[tb]
  \centering
  \includegraphics[width=\linewidth]{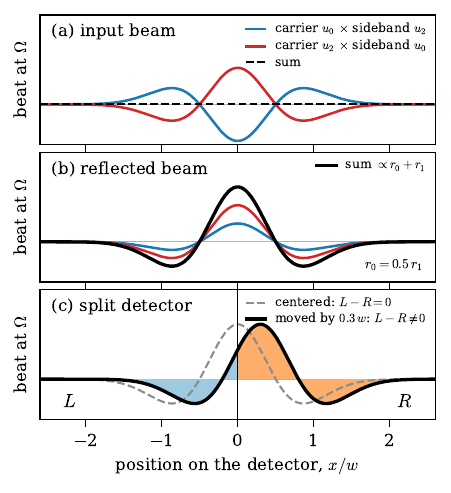}
  \caption{Beats at the frequency $\Omega$ between $u_0$ and $u_2$ across one
    transverse plane, for $\mathrm{Im}[c_2] > 0$. (a) In the input beam, the
    beat of the carrier $u_0$ with the sideband $u_2$ and that of the carrier
    $u_2$ with the sideband $u_0$ are symmetric about the beam center and
    opposite in sign, and cancel. (b) In reflection, the carrier $u_0$ returns
    with the on-resonance cavity reflectivity $r_0$ instead of the prompt
    reflection $-r_1$ of the input mirror, shown for $r_0 = 0.5\,r_1$, and a
    symmetric beat proportional to $r_0 + r_1$ remains. (c) On a detector split
    at the beam center, this beat cancels in the difference $L - R$ (dashed).
    Moving the beam by $0.3\,w$, with $w$ the beam radius, breaks the
    symmetry, and $L - R$ picks up the
    beat (solid).}
  \label{fig:beat}
\end{figure}

The carrier $u_0$ and the sideband $u_2$ beat at the frequency $\Omega$, as do the carrier $u_2$ and the sideband $u_0$, and each
beat forms a pattern symmetric about the beam center, of one sign at the
center and of the other on either side. In the input beam the two patterns have opposite signs and cancel
(Fig.~\ref{fig:beat}(a)), since the carrier and the sidebands share one
transverse profile.

The cavity breaks this cancellation. The carrier $u_0$ is resonant and
returns with the reflectivity $r_0$ of the cavity on resonance, whereas the
carrier $u_2$ and both sidebands are off resonance and are reflected promptly
by the input mirror, with $-r_1$ (Appendix~\ref{app:coeff}). The two patterns
then no longer cancel, and a symmetric beat proportional to $r_0 + r_1$ remains in the reflected light
(Fig.~\ref{fig:beat}(b)). At critical coupling ($r_0 = 0$), for instance, the carrier
$u_0$ is absent from the reflection port, and only the pattern of the carrier
$u_2$ with the sideband $u_0$ remains.

The remaining beat is, however, invisible to an ordinary photodetector, since
its integral over
the beam vanishes by the orthogonality of $u_0$ and $u_2$. A quadrant
photodetector does not see it either while the beam is centered, because the
pattern is symmetric and cancels in the quadrant differences. Moving the beam
off center breaks the symmetry and lets the quadrant differences pick up the
beat (Fig.~\ref{fig:beat}(c)). Such a motion, of the detector or of a steering mirror after the
cavity, adds a first-order amplitude $a_1'$, defined like $a_1$, and the WFS then read, after demodulation at $\Omega$,
\begin{equation}
\begin{split}
  S_{\mathrm{RF}}^{(h)} = {}& -\sqrt{2}\, G P_0\, \beta r_1\, Q_{01}\,
  (r_0 + r_1) \\
  & \times \mathrm{Im}\!\left[ c_2\, e^{-2i\eta_h} \right]
  \mathrm{Re}\!\left[ a_1'\, e^{-i\eta_h} \right]
\end{split}
  \label{eq:null}
\end{equation}
for sensor $h$ ($h = A, B$), where $\eta_h$ is the Gouy phase accumulated
from the cavity waist to the sensor plane, $G$ is the total gain of the signal
chain, $P_0$ the input power, $\beta$ the modulation index
(Appendix~\ref{app:coeff}), and $Q_{01}$ the overlap of $u_0$ and $u_1$ on
the quadrant mask, which takes the difference of the two halves of the
detector, $x > 0$ and $x < 0$ in the transverse coordinate $x$,
\begin{equation}
  Q_{01} = \int_{-\infty}^{\infty} \mathrm{sgn}(x)\, u_0(x)\, u_1(x)\, dx
  = \sqrt{\frac{2}{\pi}} .
  \label{eq:q01}
\end{equation}
The signal vanishes for a matched beam, so the
WFS read it as a null measurement of the mismatch.

Each sensor reads only one quadrature of the complex amplitude,
$\mathrm{Im}[c_2\, e^{-2i\eta_h}]$. Since the second-order mode rotates at
twice the Gouy phase, this quadrature is set by the location of the sensor,
and two sensors at different Gouy phases measure both degrees of freedom of
$c_2$, once their Gouy phases are known. The prefactor, which holds the gain
of the signal chain, the power, the modulation index, and the reflectivities,
has to be known as well, and so does the drive amplitude $|a_1'|$ of the beam
motion.

Calibrating the signal of Eq.~(\ref{eq:null}) thus requires many parameters
to be estimated. This is because the two quadratures of $c_2$ are read by two
different sensors, whose signals have to be compared in absolute value, each
with its own gain and Gouy phase. The method resolves this by
adding a reference field that the two sensors share.

\subsection{Calibration by a longitudinal dither}
\label{sec:wfs}

The same problem arises in alignment sensing, the task for which the WFS were
built. For a misalignment $a_1$ of the input beam to the
cavity, sensor $h$ reads
\begin{equation}
  S_{\mathrm{RF}}^{(h)} = G P_0\, \beta r_1\, Q_{01}\, (r_0 + r_1)\,
  \mathrm{Im}\!\left[ a_1\, e^{-i\eta_h} \right]
  \label{eq:wfs}
\end{equation}
with the symbols of Eq.~(\ref{eq:null}). Each sensor again reads one
quadrature, now of $a_1$, set by its Gouy phase and scaled by the same
prefactor, so recovering the complex $a_1$ also requires comparing the two
sensors in absolute value. The mismatch signal of Eq.~(\ref{eq:null}) and the
misalignment signal of Eq.~(\ref{eq:wfs}) come from the same sensors and share
their prefactor and Gouy phases, so calibrating the misalignment signal is
equivalent to calibrating the mismatch signal. This section therefore shows
how the reference field calibrates the misalignment signal.

The shared reference field comes from a longitudinal dither, at frequency $\omega_L$
and depth $\chi_a$ in units of the cavity half linewidth, applied to a longitudinal actuator
of the cavity, such as a length actuator or the laser frequency actuator. It phase-modulates the intracavity light and puts audio
sidebands at $\pm\omega_L$ on $u_0$. These audio sidebands beat
with the RF sidebands and add a signal at $\omega_L$ to the demodulated WFS
output. Since the dither is a phase modulation,
the audio sidebands are in quadrature with the carrier, so this beat and the
carrier beat of Eq.~(\ref{eq:wfs}) read the two orthogonal quadratures of the
same field through the one signal chain. Demodulating again at $\omega_L$
gives the double-demodulated (DD) signal, which for a misalignment $a_1$ alone
reads
\begin{equation}
  S_{\mathrm{DD}}^{(h)} = -\, G P_0\, \beta r_1\, (r_0 + r_1)\, \chi_a\,
  Q_{01}\, \mathrm{Re}\!\left[ a_1\, e^{-i\eta_h} \right],
  \label{eq:ddx}
\end{equation}
the real part of the quantity whose imaginary part the RF signal of
Eq.~(\ref{eq:wfs}) reads.

The direct-current (DC) signal of the quadrant photodetector also reads this
real part, but it
passes through a path separate from the RF chain and so cannot calibrate the
RF signal. The DD signal, in contrast, is taken from the RF-demodulated output
itself and passes through the same photodiode, resonant circuit, and
demodulator as the RF signal, so the prefactor $G P_0\, \beta r_1 (r_0 + r_1)$
of Eqs.~(\ref{eq:wfs}) and (\ref{eq:ddx}) is the same. The second
demodulation, the only stage the DD signal does not share with the RF
signal, runs at the low dither frequency, where it is readily done
numerically and its response is simple to correct.
Furthermore, the DD signal summed over all segments (SUM),
\begin{equation}
  S_{\mathrm{DD,SUM}} = G P_0\, \beta r_1\, (r_0 + r_1)\, \chi_a,
  \label{eq:ddsum}
\end{equation}
carries the same factors and does not depend on the beam position, so it
serves as a reference line. Normalizing by it cancels this prefactor exactly,
together with $\chi_a$ on the DD side,
\begin{align}
  \hat{S}_{\mathrm{DD}}^{(h)} &\equiv
  \frac{S_{\mathrm{DD}}^{(h)}}{S_{\mathrm{DD,SUM}}}
  = -\, Q_{01}\,
    \mathrm{Re}\!\left[ a_1\, e^{-i\eta_h} \right],
  \label{eq:ddnorm} \\
  \hat{S}_{\mathrm{RF}}^{(h)} &\equiv
  \frac{S_{\mathrm{RF}}^{(h)}}{S_{\mathrm{DD,SUM}}}
  = \frac{Q_{01}}{\chi_a}\,
    \mathrm{Im}\!\left[ a_1\, e^{-i\eta_h} \right],
  \label{eq:rfnorm}
\end{align}
leaving two orthogonal projections of $a_1 e^{-i\eta_h}$, scaled by $Q_{01}$
and, on the RF side, by $1/\chi_a$. Since $Q_{01}$ follows from the detector
geometry alone, Eq.~(\ref{eq:q01}), the dither depth $\chi_a$ is the only
quantity left to determine.

The Gouy phases are then calibrated with a deliberate misalignment $a_1$,
applied by an angular actuator before the cavity. An actuator at accumulated
Gouy phase $\eta_a$ tilts the beam at its own plane by an angle $\theta_a$,
normalized to the diffraction angle of the beam there, and so excites the
first-order mode with the complex amplitude
\begin{equation}
  a_1 = i\, \theta_a\, e^{i\eta_a},
  \label{eq:a1phase}
\end{equation}
the factor $i$ marking a tilt rather than a displacement. This excitation
appears in both normalized signals in proportion to the drive, so the ratios
between the two sensors cancel it together with $\chi_a$,
\begin{equation}
  \frac{\hat{S}_{\mathrm{DD}}^{(A)}}{\hat{S}_{\mathrm{DD}}^{(B)}}
  = \frac{\sin(\eta_A - \eta_a)}{\sin(\eta_B - \eta_a)},
  \qquad
  \frac{\hat{S}_{\mathrm{RF}}^{(A)}}{\hat{S}_{\mathrm{RF}}^{(B)}}
  = \frac{\cos(\eta_A - \eta_a)}{\cos(\eta_B - \eta_a)},
  \label{eq:headratios}
\end{equation}
and solving them gives $\eta_A - \eta_a$ and $\eta_B - \eta_a$. With two such
actuators, at $\eta_1$ and $\eta_2$, the ratios fix the three relative Gouy
phases among $\eta_A$, $\eta_B$, $\eta_1$, and $\eta_2$ completely. One gauge
freedom, a common offset of all four, remains and is fixed separately
(Sec.~\ref{sec:apparatus}). Once the Gouy phases are known, the ratio of the RF to
the DD signal within one sensor gives $\chi_a$.

\subsection{Reading $c_2$ from the normalized signals}
\label{sec:dprime}

The reference line of Eq.~(\ref{eq:ddsum}) normalizes the null signal of
Eq.~(\ref{eq:null}) in the same way, giving
\begin{equation}
  \hat{S}_{\mathrm{RF}}^{(h)}
  = -\,\frac{\sqrt{2}\, Q_{01}}{\chi_a}\,
    \mathrm{Im}\!\left[ c_2\, e^{-2i\eta_h} \right]
    \mathrm{Re}\!\left[ a_1'\, e^{-i\eta_h} \right].
  \label{eq:rfprime}
\end{equation}
The DD signal reads the same
motion as an $O(1)$ displacement,
\begin{equation}
  \hat{S}_{\mathrm{DD}}^{(h)}
  = -\,2\, Q_{01}
    \left( 1 - \frac{\mathrm{Re}[c_2\, e^{-2i\eta_h}]}{\sqrt{2}} \right)
    \mathrm{Re}\!\left[ a_1'\, e^{-i\eta_h} \right] ,
  \label{eq:ddprime}
\end{equation}
which also makes it a signal for centering the beam on the quadrant photodetector,
as Appendix~\ref{app:centering} sets out.

Exciting $a_1'$ with a single angular actuator
($a_1' = i\, \theta_{a'}\, e^{i\eta_{a'}}$) and
taking the ratios of the signals between the two sensors cancels every
parameter, as in the calibration with $a_1$. Writing
$c_2 = |c_2|\, e^{i\phi_2}$,
\begin{equation}
\begin{gathered}
  \frac{\hat{S}_{\mathrm{RF}}^{(A)}}{\hat{S}_{\mathrm{RF}}^{(B)}}
  = \frac{\sin(\phi_2 - 2\eta_A)\, \sin(\eta_A - \eta_{a'})}
         {\sin(\phi_2 - 2\eta_B)\, \sin(\eta_B - \eta_{a'})},
  \\
  \frac{\hat{S}_{\mathrm{DD}}^{(A)}}{\hat{S}_{\mathrm{DD}}^{(B)}}
  = \frac{\left(1 - |c_2| \cos(\phi_2 - 2\eta_A)/\sqrt{2}\right)
          \sin(\eta_A - \eta_{a'})}
         {\left(1 - |c_2| \cos(\phi_2 - 2\eta_B)/\sqrt{2}\right)
          \sin(\eta_B - \eta_{a'})}.
\end{gathered}
  \label{eq:rfprimeratio}
\end{equation}
Neither the drive amplitude, nor $\chi_a$, nor even $|c_2|$ appears in the
RF ratio, so once the Gouy phases are known it yields the phase $\phi_2$ of
$c_2$ directly. The DD ratio, which at zeroth order in $c_2$ reduces to the
geometric sine ratio of Eq.~(\ref{eq:headratios}), determines $\eta_{a'}$.

When layout constraints prevent driving a common $a_1'$ for both sensors, the
ratio within a single sensor serves instead, and the drive amplitude cancels,
\begin{equation}
\begin{split}
  \frac{\hat{S}_{\mathrm{RF}}^{(h)}}{\hat{S}_{\mathrm{DD}}^{(h)}}
  &= \frac{1}{\sqrt{2}\, \chi_a}\,
    \frac{\mathrm{Im}[c_2\, e^{-2i\eta_h}]}{1 - \mathrm{Re}[c_2\, e^{-2i\eta_h}]/\sqrt{2}} \\
  &\simeq \frac{1}{\sqrt{2}\, \chi_a}\,
    \mathrm{Im}\!\left[ c_2\, e^{-2i\eta_h} \right],
\end{split}
  \label{eq:rho}
\end{equation}
giving the projection of $c_2$ without knowing the drive amplitude. The
remaining parameters, the dither depth $\chi_a$ and the Gouy phase $\eta_h$ that
names the axis of the projection, both follow from the calibration of
Sec.~\ref{sec:wfs}.

\subsection{Full reconstruction of the mode mismatch}
\label{sec:full}

Applied to the horizontal and the vertical plane, the reading of
Sec.~\ref{sec:dprime} gives the mismatch in each direction, $c_{20}$ and
$c_{02}$. The remaining amplitude, $c_{11}$, describes astigmatism along the
diagonal axes, and the same sensors read it with no excitation at all.

With the two-dimensional modes written $u_{mn} = u_m(x)\, u_n(y)$, the beat
of $u_{11}$ with $u_0$, left uncanceled by the cavity in the same
way as that of $u_2$ in Sec.~\ref{sec:shared}, forms a pattern odd in both $x$
and $y$. The diagonal combination of the quadrants, the sum of one diagonal
pair of segments minus the sum of the other (DIAG), reads it without any beam
motion. For an aligned beam no other mode pair up to second order is odd in
both axes. Normalized by the reference line, the static DIAG signal of
sensor $h$ is
\begin{equation}
  \hat{S}_{\mathrm{DIAG}}^{(h)}
  = \frac{Q_{01}^{2}}{\chi_a}\,
    \mathrm{Im}\!\left[ c_{11}\, e^{-i\eta_h^{(11)}} \right],
  \label{eq:diag}
\end{equation}
where $\eta_h^{(11)} = \eta_{h,x} + \eta_{h,y}$ is the sum of the Gouy phases of the
two planes, with which $u_{11}$ rotates. With these phases known from
Sec.~\ref{sec:wfs}, the static values of the two sensors are two projections
of $c_{11}$, and a two-point inversion gives the complex $c_{11}$ with no
further calibration.

\section{Demonstration}
\label{sec:measured}

\subsection{Apparatus}
\label{sec:apparatus}

\begin{figure*}[htbp]
  \centering
    \includegraphics[width=0.85\textwidth]{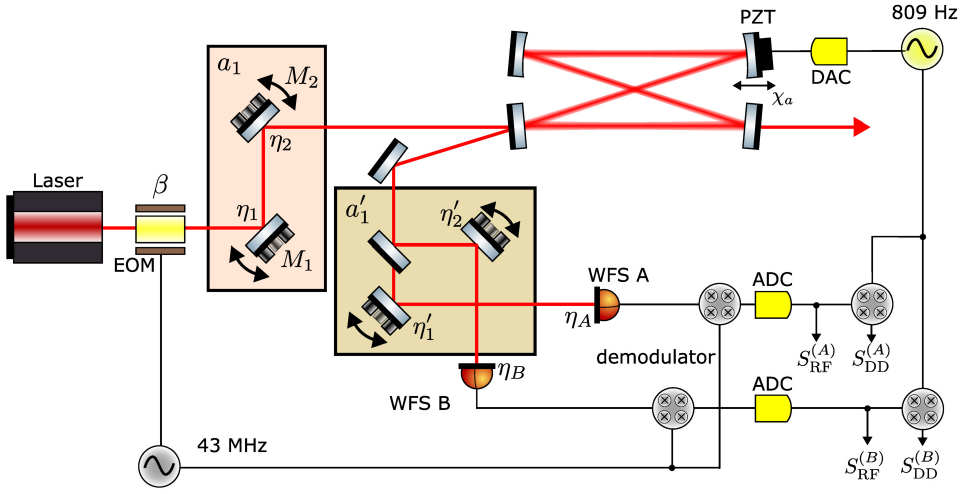}
  \caption{Layout of the demonstration. The beam is phase-modulated at
    $43$\,MHz with modulation index $\beta$ and injected into the
    jitter attenuation cavity, a four-mirror bow-tie ring cavity whose length is
    dithered at $809$\,Hz through its own piezoelectric transducer (PZT),
    giving $\chi_a$. On the input
    path, the steering mirrors M$_1$ and M$_2$, at Gouy phases $\eta_1$
    and $\eta_2$, carry the angular dithers $a_1$, applied through the
    PZT of M$_1$ and the suspension of M$_2$; on the detection path a mirror before each
    sensor, at $\eta'_1$ (WFS~A) and $\eta'_2$ (WFS~B), carries $a_1'$, applied as
    open-loop picomotor stepping. Two wavefront sensors sit at Gouy phases $\eta_A$ and
    $\eta_B$. Demodulation at $43$\,MHz forms the RF signals
    $S^{(A)}_{\mathrm{RF}}$ and $S^{(B)}_{\mathrm{RF}}$; a second,
    digital demodulation at the length-dither frequency, downstream of the
    analog-to-digital converters, forms the double-demodulated signals $S^{(A)}_{\mathrm{DD}}$ and
    $S^{(B)}_{\mathrm{DD}}$.}
  \label{fig:setup}
\end{figure*}

The demonstration uses the jitter attenuation cavity, a bow-tie ring cavity
newly installed to suppress beam jitter at the LIGO Hanford Observatory, in a
vacuum chamber between the pre-stabilized laser system~\cite{kwee2012} and the
input mode cleaner~\cite{mueller2016}; the layout is shown in
Fig.~\ref{fig:setup} and the cavity parameters in Table~\ref{tab:cavity}
(Appendix~\ref{app:cavity}). The
cavity length is held on resonance using the PDH signal, and measurements were
taken at $2$\,W and $62$\,W input.

The two wavefront sensors take their different Gouy phases from a telescope lens L1 upstream of them,
and each picomotor-driven mirror of the $a_1'$
drive, at Gouy phase $\eta'_1$ before WFS~A and $\eta'_2$ before WFS~B (the
$\eta_{a'}$ of Sec.~\ref{sec:dprime} for each sensor), walking the beam in
small discrete steps, sits $3$ to $5$ inches before its sensor. In the acquisition chain
the quadrant photodetector signals pass through a
resonant circuit, a demodulator, and a whitening filter to the
analog-to-digital converters of a $16$\,kHz digital
system~\cite{bork2021}, both demodulator quadratures ($I$ and $Q$) recorded
for every segment. All Gouy phases are measured from the cavity waist. Of the
input steering mirrors M$_1$ and M$_2$ (Fig.~\ref{fig:setup}), M$_2$ is the
Gouy-phase anchor (Appendix~\ref{app:design}), which fixes the common offset
of the Gouy phases (Sec.~\ref{sec:wfs}): the path between M$_2$ and the cavity
waist is pure free propagation, and its measured length ($0.56$\,m) is
converted into the Gouy phase of M$_2$.

\subsection{Signal acquisition and processing}
\label{sec:acq}

The measurement uses two families of records. One family carries the four
angular calibration lines for the Gouy phases (Sec.~\ref{sec:wfs}).
They tilt M$_1$ and M$_2$ in yaw (YAW, horizontal) and pitch (PIT, vertical),
all four
driven simultaneously in a single record in the $2$\,W resonant state (duration
$7200$\,s), each at a distinct frequency ($9.7$ to $14.9$\,Hz).
The other family carries the picomotor sweeps for the $a_1'$ excitation that reads
$c_2$ (Sec.~\ref{sec:dprime}), two sets
at $62$\,W (runs 1 and 2) and one set at $2$\,W (with the alignment control
running), each set comprising four records (both sensors $\times$ two planes).
The picomotor was swept back and
forth across the operating point in a triangle wave (peak-to-peak $50$ steps,
fundamental frequency $\simeq 3.95$\,Hz). The $809$\,Hz length dither was
applied continuously in all records.

The segments are combined into PIT (the upper pair of segments minus the
lower pair), YAW (the left pair minus the right pair), and DIAG, with YAW
reading $c_{20}$ and PIT $c_{02}$. These combinations are re-demodulated at
the dither frequency to form the DD signals, and both the RF and the DD
signals are normalized by $S_{\mathrm{DD,SUM}}$
[Eq.~(\ref{eq:ddsum})]. The line and sweep amplitudes are taken by coherent
projection in blocks, the block-to-block scatter giving the statistical
error, and an unmeasured systematic error of $5$\,\% of the value is added in
quadrature for the line observables.

\begin{figure}[htbp]
  \centering
  \includegraphics[width=\linewidth]{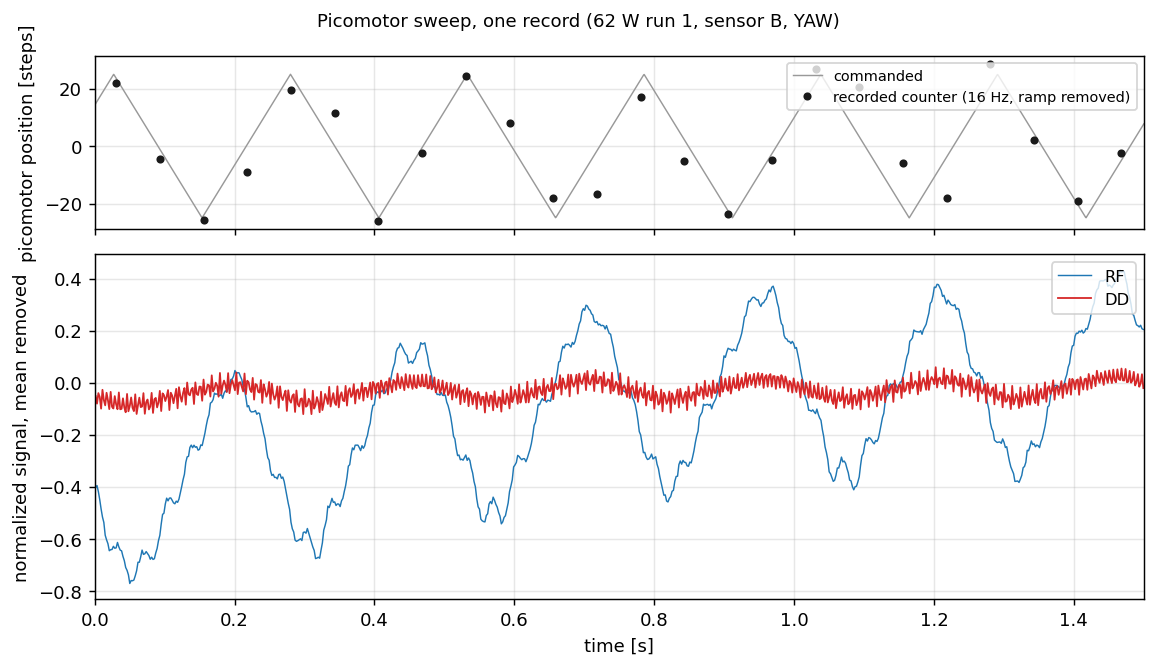}
  \caption{One record of the picomotor sweep ($62$\,W run 1, sensor B, YAW).
    Top: the picomotor position, the recorded counter (the linear ramp of the
    driver's step-efficiency compensation removed) together with the
    commanded triangle. Bottom: the SUM-normalized DD and RF signals (mean
    removed). The dither depth of this record is $\chi_a \simeq 0.004$.}
  \label{fig:picots}
\end{figure}

Figure~\ref{fig:picots} shows one record of the picomotor sweep. Both normalized
signals follow the triangular motion, the DD signal with the geometric
coefficient of Eq.~(\ref{eq:ddprime}) and the RF signal only through $c_2$,
amplified by $1/\chi_a$ [Eq.~(\ref{eq:rfprime})], and the ratio of the two traces
is the observable of Eq.~(\ref{eq:rho}).

\subsection{Joint fit}
\label{sec:fit}

At zeroth order in $c_2$, the $a_1$ ratios give the Gouy phases
[Eq.~(\ref{eq:headratios})] and the $a_1'$ ratios give $c_2$
[Eqs.~(\ref{eq:rfprimeratio}) and (\ref{eq:rho})]. Each of these expressions,
however, carries first-order terms in $c_2$ that are in general not
negligible. The second-order mode contaminates the misalignment signals of
Eqs.~(\ref{eq:ddnorm}) and (\ref{eq:rfnorm}) by $10$ to $20$\,\% for a
mismatch of a few percent in power (Appendix~\ref{app:coeff2}), so the
analysis treats all the ratios as one system of equations and determines the
Gouy phases and $c_2$ simultaneously in a single least-squares fit. The contamination is a known function of $c_2$ and of the Gouy phase of each
sensor (Appendix~\ref{app:coeff2}), so the joint fit reads it as a further
measurement of $c_2$ instead of treating it as an error.

Each run, the line record together with one set of picomotor sweeps, is
analyzed in an independent least-squares fit, with ten observables in each plane. Eight are
the normalized RF and DD signals of the two calibration lines of that plane,
the tilts of M$_1$ and M$_2$, each read at both sensors
[Eqs.~(\ref{eq:ddnorm}) and (\ref{eq:rfnorm})]. The other two are the RF/DD ratios
of the picomotor sweep, one at each sensor [Eq.~(\ref{eq:rho})]. The parameters are,
in each plane, the Gouy phases of the two sensors and of M$_1$, with that of
M$_2$ fixed from its measured distance as the anchor
(Sec.~\ref{sec:apparatus}), and the complex $c_{20}$ or
$c_{02}$, together with five scale parameters in all, one $k$ per calibration line and
one $k_\chi$ for the dither depth, each the ratio of the true excitation to an
independent estimate from the transmitted power (Appendix~\ref{app:dither}).
These scales calibrate the actuators and the dither as a by-product
(Sec.~\ref{sec:actcal}). The model carries $c_2$ to second order (Appendix~\ref{app:coeff2}).
Calibration lines were not applied at $62$\,W, and the $62$\,W fits use the
line record taken at $2$\,W in their place (Appendix~\ref{app:proc}).

\subsection{Results}

The fit quality, $\chi^2$ per degree of freedom (dof), is
$\chi^2/\mathrm{dof} = 2.7/9 = 0.30$ at $2$\,W and
$1.34/5 = 0.27$ for both $62$\,W runs. The results are discussed further in
Sec.~\ref{sec:discussion}.

\begin{figure*}[htbp]
  \centering
  \includegraphics[width=\linewidth]{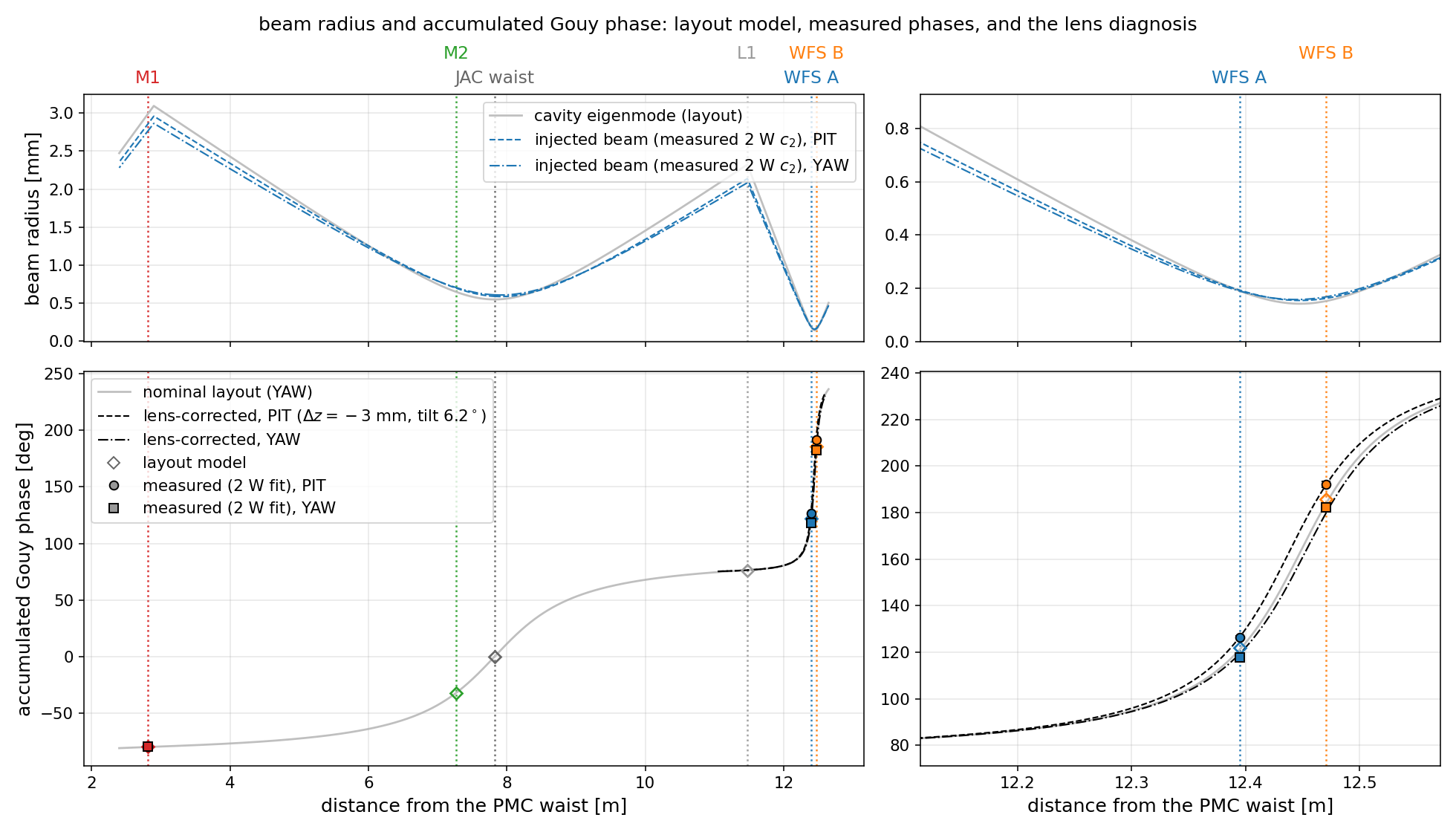}
  \caption{Beam radius (top) and accumulated Gouy phase (bottom) along the
    path, against the distance from the waist of the pre-mode cleaner (PMC).
    The Gouy phase is measured
    from the cavity waist, and M$_2$ is the anchor that fixes the common
    offset of the measured Gouy phases (Sec.~\ref{sec:apparatus}). The gray curves and open markers
    are the layout model; the blue curves in the top row are the injected
    beam predicted from the measured $2$\,W $c_2$ (dashed PIT, dash-dotted
    YAW); the filled markers are the
    measured values ($2$\,W fit), with error bars smaller than the markers;
    the black dashed and dash-dotted curves
    are the per-plane model corrected by two telescope-lens mounting errors
    ($\Delta z = -3$\,mm, tilt $6.2^\circ$; Sec.~\ref{sec:geom}).
    The right column zooms in on the sensor region: at the telescope focus the
    beam is small and the phase accumulates steeply. JAC: jitter attenuation
    cavity, the cavity under test.}
  \label{fig:layoutgouy}
\end{figure*}

\medskip
\paragraph{Gouy phases.}
Figure~\ref{fig:layoutgouy} and Table~\ref{tab:gouy} compare the fitted Gouy
phases with an optical-layout model.

The Gouy phase of M$_1$ lands on the
layout model's $280.01^\circ$ (modulo $360^\circ$; $-79.99^\circ$ in
Fig.~\ref{fig:layoutgouy}) to within $0.3^\circ$, with a split between the
planes of $0.10^\circ$. The sensor separation $\eta_A - \eta_B$ is
$-65.4^\circ$ in PIT and $-64.3^\circ$ in YAW; the two planes agree to
$1.1^\circ$. The Gouy phases of the two sensors deviate from the model by $4$ to $9^\circ$,
in opposite directions in the two planes.

\medskip
\paragraph{Scale parameters.}
All five scales of Table~\ref{tab:scales} sit below unity, by $2$ to
$4$\,\%.

\medskip
\paragraph{Complex amplitudes of the second-order modes.}
Table~\ref{tab:c2} collects the per-run $c_2$ and the $c_{11}$ from the
static DIAG.

The two $62$\,W runs agree within errors, and $c_{20}$ grows significantly
from its $2$\,W value. In contrast, $c_{11}$ is nearly independent of power.

\begin{table}[H]
\centering
\caption{Fitted Gouy phases ($2$\,W fit) in degrees, measured from the
cavity waist and given modulo $360^\circ$ (M$_1$ appears at $-80^\circ$ in
Fig.~\ref{fig:layoutgouy}), with M$_2$ as the anchor that fixes their common
offset (Sec.~\ref{sec:apparatus}). Parentheses give the error in units of the last digit, square brackets the
layout-model value.}
\label{tab:gouy}
\begin{tabular}{lcc}
  \toprule
  & PIT & YAW \\
  \midrule
  WFS A & $126.4(6)$ \;[$120.2$] & $117.7(7)$ \;[$121.9$] \\
  WFS B & $191.8(12)$ \;[$182.4$] & $182.0(11)$ \;[$185.7$] \\
  M$_1$ & $279.9(12)$ \;[$280.0$] & $279.8(10)$ \;[$280.0$] \\
  \bottomrule
\end{tabular}
\end{table}

\begin{table}[H]
\centering
\caption{The five scale parameters of the $2$\,W fit, each the ratio of the
measured quantity to its independent estimate.}
\label{tab:scales}
\begin{tabular}{lc}
  \toprule
  Scale & $2$\,W \\
  \midrule
  $k$ (M$_1$, PIT) & $0.967(17)$ \\
  $k$ (M$_1$, YAW) & $0.981(18)$ \\
  $k$ (M$_2$, PIT) & $0.956(17)$ \\
  $k$ (M$_2$, YAW) & $0.962(15)$ \\
  $k_\chi$ (dither depth) & $0.962(15)$ \\
  \bottomrule
\end{tabular}
\end{table}

\begin{table}[H]
\centering
\caption{Measured complex amplitudes of the second-order modes (phases
relative to $u_0$ at the cavity waist): amplitude, phase, and power $|c|^2$.
The two $62$\,W runs are separate records that share the geometry of the
$2$\,W line record, and $c_{11}$ at $62$\,W is one value for both. Powers
are computed from the unrounded amplitudes, and their errors are propagated
from the amplitude errors as
$\sigma_{|c|^2} = 2|c|\,\sigma_{|c|}$, added in quadrature across modes for
the total-power row. The $c_{11}$ is corrected for the light lost in the gaps
between the quadrant segments (Sec.~\ref{sec:systematics}).}
\label{tab:c2}
\small
\setlength{\tabcolsep}{4pt}
\begin{tabular}{llccc}
  \toprule
  & & $2$\,W & $62$\,W run 1 & $62$\,W run 2 \\
  \midrule
  $c_{20}$ & amplitude & $0.084(7)$ & $0.118(5)$ & $0.121(5)$ \\
           & phase & $+26(2)^\circ$ & $+38(2)^\circ$ & $+35(2)^\circ$ \\
           & power & $0.70(11)\,\%$ & $1.39(12)\,\%$ & $1.46(12)\,\%$ \\
  \midrule
  $c_{02}$ & amplitude & $0.062(7)$ & $0.057(4)$ & $0.062(3)$ \\
           & phase & $+38(2)^\circ$ & $+46(3)^\circ$ & $+42(3)^\circ$ \\
           & power & $0.38(9)\,\%$ & $0.33(4)\,\%$ & $0.38(4)\,\%$ \\
  \midrule
  $c_{11}$ & amplitude & $0.045(3)$ & \multicolumn{2}{c}{$0.045(3)$} \\
           & phase & $+57(3)^\circ$ & \multicolumn{2}{c}{$+53(3)^\circ$} \\
           & power & $0.20(2)\,\%$ & \multicolumn{2}{c}{$0.21(2)\,\%$} \\
  \midrule
  \multicolumn{2}{l}{total power $\sum|c|^2$} & $1.3(1)\,\%$ & $1.9(1)\,\%$ & $2.0(1)\,\%$ \\
  \bottomrule
\end{tabular}
\end{table}

\newpage
\section{Discussion}
\label{sec:discussion}

\subsection{Comparison with the cavity scan}
\label{sec:cross}

We compare the results with an independent cavity scan, taken with the cavity
out of resonance control and so carrying essentially no thermal load. The
cavity length is swept over one free spectral range, and the second-order mode
content follows from the heights of the second-order peaks in the transmitted
power. The scan returns a second-order mode content of $1.15\,\%$ at $2$\,W
and $1.0\,\%$ and $1.5\,\%$ in two repetitions at $62$\,W. The two $62$\,W scans put the repeat-to-repeat scatter at
$\pm 0.25\,\%$, which we adopt as the systematic error of a scan at both
powers; against it the scan shows no significant difference between $2$\,W and
$62$\,W.

At $2$\,W, where the thermal load is small and both measurements see the
same eigenmode, our method agrees with the scan, returning $1.28\,\%$ in total,
above the scan by $0.13\,\%$ and within the systematic error. At $62$\,W it returns
$1.9$ to $2.0\,\%$ (Table~\ref{tab:c2}), above the $62$\,W scans' average of
$1.25\,\%$ by about three times the systematic error. This departure appears only under thermal load, and its
size is the thermal change of the eigenmode that the resonant measurement
follows and the scan cannot reach.

\subsection{Actuator calibration}
\label{sec:actcal}

The scale parameters of Table~\ref{tab:scales} compare the line excitations
and the dither depth that the fit measures with their independent estimates
from the transmitted power (Appendix~\ref{app:dither}), and they sit $2$ to
$4$\,\% below unity. Two corrections apply. The transmitted-power estimates
read $3.9$\,\% high through the digital whitening compensation of their chain
(Appendix~\ref{app:dither}), and the fit reads the line excitations $6$ to
$9$\,\% high through the gap between the quadrant segments
(Sec.~\ref{sec:systematics}), which leaves the dither depth untouched. With
both corrections, the dither-depth scale becomes $1.00$ and the four line
scales $0.91$ to $0.96$. One possible origin of this remaining difference is
the coupling of each angular actuator between pitch and yaw. The transmitted
power responds to the motion in both planes, while the fit reads each plane
separately, so a motion in the other plane raises the estimate but not the
fitted excitation. The dither depth, which has no such coupling, agrees with
its estimate. The difference leaves $c_2$ almost untouched, since $c_2$ is
read from ratios of signals in which the scales cancel.

\subsection{Reconstruction of the optical geometry}
\label{sec:geom}

\begin{figure*}[htbp]
  \centering
  \includegraphics[width=\linewidth]{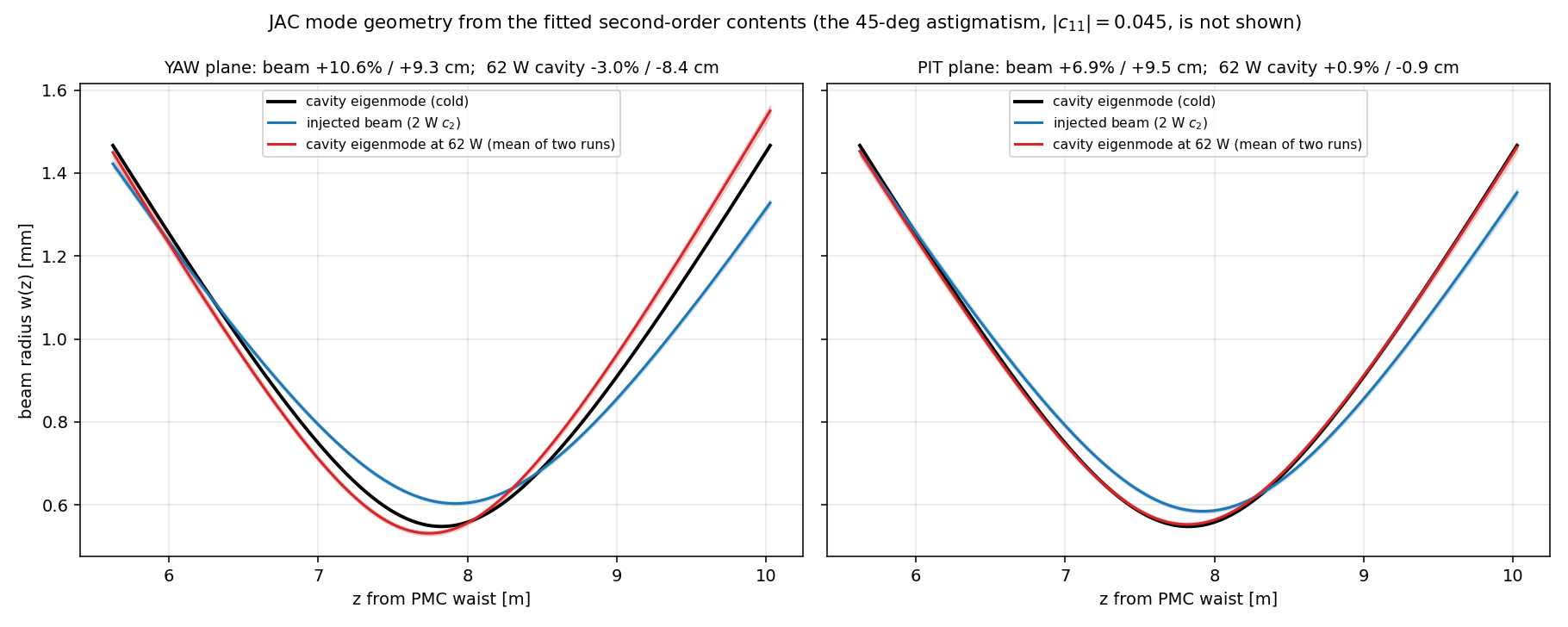}
  \caption{Beam geometry reconstructed from the measured second-order modes.
    Black: the cold cavity eigenmode; blue: the input beam (from the $2$\,W
    $c_2$); red: the $62$\,W eigenmode (from the difference of the $c_2$, with the mean
    of the two $62$\,W runs).
    Bands are $\pm 1\sigma$. The panel titles quote waist-size offsets as
    fractions of the cold waist radius and waist-position offsets in
    centimeters. PMC: pre-mode cleaner; JAC: jitter attenuation cavity, the
    cavity under test.}
  \label{fig:mmgeom}
\end{figure*}

Through the standard second-order overlap of two Gaussian
beams~\cite{anderson1984,bayerhelms1984}, the complex amplitudes convert into
a waist-size offset $\delta w_0$ and a waist-position offset $\delta z$ of the
input beam relative to the cavity eigenmode,
\begin{equation}
  \frac{\delta w_0}{w_0} = \sqrt{2}\,\mathrm{Re}[c_2],
  \qquad
  \delta z = 2\sqrt{2}\, z_R\, \mathrm{Im}[c_2],
  \label{eq:geom}
\end{equation}
where $w_0$ is the waist radius and $z_R$ the Rayleigh range. The scan shows no significant change between the two powers
(Sec.~\ref{sec:cross}), so the input side is taken as fixed, and the change of
$c_2$ from $2$\,W to $62$\,W is read as a change of the cavity eigenmode at
a fixed input beam. Both sets of offsets are collected in
Table~\ref{tab:geomoffsets}. The thermal eigenmode change is concentrated in
the horizontal plane, with the waist displacement in units of $z_R$ about
three times the size reduction in units of $w_0$ (the two $62$\,W runs are
consistent).
Figure~\ref{fig:mmgeom} shows the beam geometry reconstructed from the
measured $c_2$.

\begin{table}[htbp]
\centering
\caption{Mode mismatch converted into beam offsets with
Eq.~(\ref{eq:geom}). The input-beam mismatch is the offset of the input beam
from the cavity eigenmode in the $2$\,W fit; the thermal change is the change
of the eigenmode from $2$\,W to the mean of the two $62$\,W runs at a fixed input
beam. Size offsets are given as ratios to the waist radius, $\delta w_0/w_0$
($w_0 \simeq 548$\,\textmu m), and position offsets as ratios to the Rayleigh
range, $\delta z/z_R$ ($z_R \simeq 0.89$\,m). Errors are
propagated from the fitted $c_2$, with the $2$\,W and $62$\,W fits taken as
independent.}
\label{tab:geomoffsets}
\begin{tabular}{lcc}
  \toprule
   & Horizontal & Vertical \\
  \midrule
  \multicolumn{3}{l}{Input-beam mismatch ($2$\,W)} \\
  \quad waist size, $\delta w_0/w_0$     & $+0.106(8)$ & $+0.069(8)$ \\
  \quad waist position, $\delta z/z_R$ & $+0.10(1)$ & $+0.11(1)$ \\
  \midrule
  \multicolumn{3}{l}{Thermal change of the eigenmode ($2 \to 62$\,W)} \\
  \quad waist size, $\delta w_0/w_0$     & $-0.030(9)$ & $+0.009(8)$ \\
  \quad waist position, $\delta z/z_R$ & $-0.10(1)$ & $-0.01(2)$ \\
  \bottomrule
\end{tabular}
\end{table}

That $c_{11}$ is nearly independent of power (Table~\ref{tab:c2}) reinforces
this picture: the astigmatism produced by the thermal deformation is
essentially aligned with the detection coordinates and generates no new
$45^\circ$ component; the change appears entirely on the
$c_{20}$/$c_{02}$ side.

The reconstructed geometry can also guide a correction. Moving each input-path lens along the beam
in the model shows that a single lens displaced by about $+5$\,cm reduces the total
mismatch of both planes to $0.11$\,\% at $2$\,W and $0.26$\,\% at $62$\,W,
the remainder being a plane-asymmetric component that no single spherical
lens can address. This correction has not yet been applied to the setup,
owing to time constraints.

The deviations of the sensor Gouy phases from the layout model
(Table~\ref{tab:gouy}) are explained by the telescope lens L1 just before
the sensors. Displacing it by $\Delta z = -3$\,mm and tilting it by $6.2^\circ$
about the horizontal axis, a tilt that makes its focal length $1.2$\,\%
shorter in the vertical plane than in the horizontal, leaves all four
residuals within $2.2\sigma$ (black dashed and dash-dotted curves in
Fig.~\ref{fig:layoutgouy}).

\subsection{Systematic errors}
\label{sec:systematics}

The dominant uncertainty is the systematic error assumed for the line observables,
an unmeasured $5$\,\% of the value (Sec.~\ref{sec:acq}). When it is reduced,
the $\chi^2/\mathrm{dof}$ of every run crosses unity near $2.5$\,\%, so the
data support about half the assumption and the quoted results are
conservative, without inflated uncertainties, since the covariance is scaled
by $\chi^2/\mathrm{dof}$. What follows asks where this $2.5$\,\% floor comes from.

Truncating the model at first order in $c_2$ shifts $\mathrm{Im}[c_{20}]$ at
$62$\,W by $20$\,\%, nearly three times its statistical error, so the fits
carry the second order. The third order is expected at the level of the floor
and can be added by the same route; the other second-order products, of $c_2$
with the static miscentering, the drives, or the detuning, stay below the
floor, and the $\chi^2$ floor did not move when the second order was added,
so the expansion does not set the $2.5$\,\% (Appendix~\ref{app:coeff2}).

The sensors are modeled as ideal quadrants, but a real sensor has gains that
differ from segment to segment and a gap between its segments, and both
produce errors. First, the gains. A gain difference between the segments adds
a static offset, unread here, that reaches the line observables only through
its small product with the static miscentering. What remains is the leak
between the horizontal and the vertical plane: the gain difference makes a
sensor read part of a motion in the other plane as motion in its own. Each
calibration line is driven at its own frequency, so the intended motions of
the two planes do not mix, but each actuator also moves the beam in the other
plane through its mechanical cross-coupling, at the same frequency, and this
unintended motion leaks into the observable. With a $5$\,\% gain difference,
typical of four independent RF chains, a $10$ to $15$\,\% mechanical
cross-coupling enters the observable at $0.5$ to $0.8$\,\%.

Next, the gap between the segments. It changes $Q_{01}$, Eq.~(\ref{eq:q01}),
only at second order in its width, since the mask changes sign where the
integrand vanishes. The factor $Q_{01}$ cancels in the ratios between signals, so $c_2$
and the Gouy phases are free of it. The reference line
$S_{\mathrm{DD,SUM}}$, which collects the whole beam, instead loses the light
falling on the gap at first order. Again, this loss is a scale common to all
the normalized signals, so it cancels in the ratios that give $c_2$ and the
Gouy phases, and of the mismatch amplitudes only $c_{11}$, read in absolute
value through Eq.~(\ref{eq:diag}), is affected. The total DC photocurrent of the four segments drops by $18$\,\% (WFS~A)
and $12$\,\% (WFS~B) when the beam moves from a single segment, where it is
parked away from the gap, to its operating position. This
raises the line scales by about $6$ to $9$\,\% (Sec.~\ref{sec:actcal}) and the
reading of $c_{11}$ by $13$ to $20$\,\% depending on the sensor. The $c_{11}$ of
Table~\ref{tab:c2} is corrected for it, assuming that the RF chain loses the
same fraction as the DC photocurrent; the correction lowers its amplitude by
$12$\,\% and moves its phase by less than $1^\circ$.

Besides the sensors, the astigmatism $c_{11}$ also couples the two planes,
since a beam motion in one plane turns $u_{11}$ into a first-order mode in
the other, with a signal about $4.5$\,\% of the in-plane response, as
$|c_{11}| \simeq 0.045$. The unintended motion from the mechanical
cross-coupling of the actuators enters the observable directly in this way, at
$0.45$ to $0.7$\,\%, and the intended motion returns to its own plane through
the gain difference at $0.2$\,\%.

A residual detuning $\chi_0$ makes the reflection of the resonant carrier
$u_0$ complex (Appendix~\ref{app:coeff}), and its imaginary part
adds to Eq.~(\ref{eq:rfprime}) a term in $\chi_0\,\mathrm{Re}[a_1'
e^{-i\eta_h}]$ of the same form as the $c_2$ term, so an unmodeled detuning
reads as an offset on the projection of $c_2$. The model therefore takes the
measured $\chi_0$ as a known input (Appendix~\ref{app:proc}). The detuning
drifts during a record while the model carries its static value, but in the
model of the calibration lines it multiplies only the term in
$\mathrm{Re}[a_1 e^{-i\eta_h}]$ (Appendix~\ref{app:coeff}), itself half a
percent of the leading one, so a drift comparable to its value contributes about $0.2$\,\%.

No named term reaches $2.5$\,\% on its own. The largest sit below $1$\,\%
each and together reach $1.4$ to $1.9$\,\%, most of the floor, while nothing in
the residuals points to a mechanism of a different kind, and $2.5$\,\% is
ample for a measurement of mode mismatch. Deterministic biases, which leave
$\chi^2$ unchanged, are not counted in that floor (the truncation shift of
Table~\ref{tab:bias} is one), and two others act only on the phase. The anchor distance ($\pm 0.5$ inch, $\pm 0.6^\circ$ in angle) rotates
the phase of $c_2$ uniformly, leaving run-to-run comparisons intact. The
thermal change of the eigenmode moves the sensor-plane Gouy phases by at most
a few degrees, and since the $62$\,W fits take their geometry from the $2$\,W
record (Sec.~\ref{sec:fit}), it chiefly rotates the phase of the $62$\,W
$c_2$, by about the quoted phase errors, and reaches the eigenmode-change
decomposition (Sec.~\ref{sec:geom}) through the difference of the two states.
Neither touches the amplitudes or powers, and actuator drive efficiencies and
beam sizes do not enter the results at all.

The model also takes every off-resonant field, the carrier $u_2$ and the RF
sidebands, to be reflected promptly by the input mirror with $-r_1$. In the
cavity these fields lie at finite distances from their own resonances, so
their reflectivities depart from $-r_1$: the upper and the lower RF sideband
of the second-order mode are reflected differently, and the higher-order modes
differently from one another. These departures are hard to evaluate in closed
form and call for a numerical model of the cavity, such as
Finesse~\cite{freise2004}; computing their effect on the results is left for
future work.

\section{Summary and outlook}
\label{sec:summary}

LDWFS measures, on a cavity in its operating
state, the Gouy phases of the sensors and actuators and the complex amplitudes
of the second-order modes, with no actuator calibration, no signal-chain
calibration, and no assumed Gouy phase apart from one common offset, which is
fixed from a measured distance. The longitudinal dither provides a reference field that the separate sensors
share, and with that field each sensor also reads the quadrature that the RF
sidebands alone cannot measure.

At $2$\,W the actuator Gouy phase agreed with the optical-layout
model to $0.3^\circ$, and the $62$\,W fits reused that geometry. The total
$2$\,W mismatch agreed with an independent
cavity scan (Sec.~\ref{sec:cross}), and the three second-order components were
measured in amplitude and phase. The two $62$\,W runs agreed within their
errors, and $c_{11}$ was nearly independent of power, while $c_{20}$ grew with
power. At $62$\,W the
thermal eigenmode change was decomposed into waist-size and waist-position
changes (Sec.~\ref{sec:geom}). The $\chi^2$ floor of the fits was at the
$2.5$\,\% level, and the named systematic terms accounted for $1.4$ to
$1.9$\,\% of it (Sec.~\ref{sec:systematics}).

The next goal is the LIGO interferometer. Its alignment sensing already places
wavefront sensors at every port, at different Gouy phases~\cite{barsotti2010},
and its length control acts on every length degree of freedom, so the method
needs no new components there. What it does need is an extension from a
single cavity to the coupled cavities of the interferometer. With that
extension, the method would give the corrective actuators, PSAMS and TCS, the
closed-loop error signals they lack today, and through better mode matching
we aim to raise the sensitivity of LIGO.

\begin{acknowledgments}

The author thanks the installation team of the jitter attenuation cavity at
the LIGO Hanford Observatory, and G.~Vajente for discussions of the
calculations behind the measurement results.

In preparing this work, the author used an AI assistant (Claude, Anthropic)
for drafting and editing the manuscript and for writing the analysis code.
All derivations, results, and conclusions were verified by the author, who
takes full responsibility for the content.

LIGO was constructed by the California Institute of Technology and
Massachusetts Institute of Technology with funding from the National
Science Foundation, and operates under cooperative agreement PHY-2309200.
This paper carries LIGO Document Number LIGO-P2600491.

\end{acknowledgments}

\appendix

\section{Derivation of the response coefficients}
\label{app:coeff}

Every response coefficient in the main text follows from one bookkeeping
scheme. In
the local basis of the detection plane, the reflected field is written as
follows. Of the carrier, only the resonant $u_0$ component receives the
reflection response $r_c$ (real $r_0$ on resonance), while every other mode
receives the prompt reflection $-r_1$:
\begin{equation}
  E_c = a_0 \left[ r_c\, u_0 - r_1 \sum_{n \geq 1} \tilde{c}_n\, u_n \right],
\end{equation}
with $a_0 = \sqrt{P_0}$. Here
$\tilde{c}_n \equiv c_n e^{-in\eta}$ is the complex amplitude of each
higher-order mode ($c_1 \equiv a_1$) rotated to the Gouy phase $\eta$ of the
detection plane, written $\tilde{c}_{n,h}$ for sensor $h$, with superscripts
$r$ and $i$ for its real and imaginary parts; for $u_{11}$, which rotates with
the sum of the Gouy phases of the two planes,
$\tilde{c}_{11} \equiv c_{11}\, e^{-i\eta_h^{(11)}}$ [Eq.~(\ref{eq:diag})].
The RF sidebands are promptly reflected in all modes,
$E_\pm = \mp a_0 \frac{\beta}{2} r_1 [\, u_0 + \sum_{n \geq 1} \tilde{c}_n u_n ]$,
and the audio sidebands created by the dither ride only on the
resonant $u_0$ carrier, with amplitude proportional to $(r_0+r_1)\,\chi_a$.
The overlap integrals of the quadrant
mask $M$, which weights each segment by $\pm 1$ according to the combination,
leave, in the PIT/YAW combinations, only pairs whose indices differ by
an odd number in one axis, with
$\langle u_0 | M | u_1 \rangle = Q_{01} = \sqrt{2/\pi}$, and in the DIAG
combination only pairs odd in both axes, with $Q_{01}^2$. The $\Omega$
component of the detector output is
$2\,\mathrm{Re}\{[\langle E_c|M|E_+\rangle
+ \langle E_-|M|E_c\rangle]\, e^{i\Omega t}\}$ multiplied by $G$. The denominator of
the normalization is the beat between the audio sidebands and the RF
sidebands; the same bookkeeping gives
$S_{\mathrm{DD,SUM}} = G P_0 \beta r_1 (r_0+r_1)\, \chi_a$.

As a worked example we derive the static DIAG value of Eq.~(\ref{eq:diag}).
For a field carrying $c_{11}$,
\begin{align}
  \langle E_c | M | E_+ \rangle
  &= -\,a_0^2 \frac{\beta}{2} r_1 Q_{01}^2
     \left[ r_c^*\, \tilde{c}_{11} - r_1\, \tilde{c}_{11}^* \right],
  \\
  \langle E_- | M | E_c \rangle
  &= \phantom{-}\,a_0^2 \frac{\beta}{2} r_1 Q_{01}^2
     \left[ -\,r_1\, \tilde{c}_{11} + r_c\, \tilde{c}_{11}^* \right],
\end{align}
and on resonance ($r_c = r_0$) the sum is
$-\,i\, \beta r_1 Q_{01}^2 P_0 (r_0 + r_1)\, \mathrm{Im}[\tilde{c}_{11}]$.
Converting to the $\Omega$ component and normalizing by
$S_{\mathrm{DD,SUM}}$,
\begin{equation}
  \hat{S}_{\mathrm{DIAG}}
  = \frac{G P_0\, \beta r_1\, Q_{01}^2 (r_0+r_1)\,
          \mathrm{Im}[\tilde{c}_{11}]}
         {G P_0\, \beta r_1 (r_0+r_1)\, \chi_a}
  = \frac{Q_{01}^2}{\chi_a}\, \mathrm{Im}[\tilde{c}_{11}],
\end{equation}
which is Eq.~(\ref{eq:diag}), the form adopted in the main text.

The other coefficients of the main text follow in the same way. The
first-order responses [Eqs.~(\ref{eq:wfs}), (\ref{eq:ddnorm}), and
(\ref{eq:rfnorm})] come from the $u_0$--$u_1$ pairs of the bookkeeping
above; the second-order contamination coefficients [Eqs.~(\ref{eq:rfcont})
and (\ref{eq:ddcont})] and the displacement responses [Eqs.~(\ref{eq:rfprime})
and (\ref{eq:ddprime})] are obtained by passing the ladder coefficients of the
displacement operator ($\sqrt{2}$ for $u_2 \to u_1$, $1$ for
$u_{11} \to u_{01}$) through the same beat bookkeeping.

The terms due to residual deviation from resonance follow from the same
bookkeeping. For a detuning $\chi$ in half-linewidth units, the carrier
reflection response becomes $r_c \simeq r_0 - i(r_0+r_1)\,\chi$, where the
dither oscillates the detuning as $\chi(t) = \chi_0 + \chi_a \sin\omega_L t$ around
the static residual $\chi_0$. There are two consequences. First, on the full aperture: since
$\mathrm{Im}[r_c] \neq 0$, the RF SUM becomes proportional to $\chi$. This
is the PDH error signal itself. Second, in the quadrant difference signals,
the imaginary part of $r_c$ turns part of the carrier beat into the other
quadrature: the normalized RF signal, which on resonance reads
$\mathrm{Im}[a_1 e^{-i\eta_h}]$, picks up $-\chi\,\mathrm{Re}[a_1 e^{-i\eta_h}]$,
the quadrature the DD signal reads. A beam motion $a_1'$ at the detection
plane brings $r_c$ into both beats, so its term doubles,
$-2\chi\,\mathrm{Re}[a_1' e^{-i\eta_h}]$. In a
well-controlled cavity $\chi_0$ is of order $10^{-2}$, an order of magnitude
below the $c_2$ terms. In the fits of this demonstration, the static
residual, obtained without calibration from the ratio of the
$1\omega_L$/$2\omega_L$ lines in the transmitted power, is supplied as a
known input, and these terms are included in the model
(Appendix~\ref{app:proc}).

\section{Second-order extension of the response coefficients}
\label{app:coeff2}

The second-order mode also contaminates the misalignment signals of
Eqs.~(\ref{eq:ddnorm}) and (\ref{eq:rfnorm}). To first order in $c_2$, the
bookkeeping of Appendix~\ref{app:coeff} gives
\begin{align}
  \hat{S}_{\mathrm{RF}}^{(h)} &= \frac{Q_{01}}{\chi_a}
    \left[ \left( 1 + \frac{\tilde{c}_{2,h}^{\,r}}{\sqrt{2}} \right)
      \mathrm{Im}\!\left[ a_1 e^{-i\eta_h} \right] \right. \nonumber \\
    &\hspace{4em} \left.
      {} - \frac{3}{\sqrt{2}}\, \tilde{c}_{2,h}^{\,i}\,
      \mathrm{Re}\!\left[ a_1 e^{-i\eta_h} \right] \right],
  \label{eq:rfcont} \\
  \hat{S}_{\mathrm{DD}}^{(h)} &= -\, Q_{01}
    \left[ \left( 1 - \frac{3}{\sqrt{2}}\, \tilde{c}_{2,h}^{\,r} \right)
      \mathrm{Re}\!\left[ a_1 e^{-i\eta_h} \right] \right. \nonumber \\
    &\hspace{4em} \left.
      {} - \frac{1}{\sqrt{2}}\, \tilde{c}_{2,h}^{\,i}\,
      \mathrm{Im}\!\left[ a_1 e^{-i\eta_h} \right] \right].
  \label{eq:ddcont}
\end{align}
Through its Gouy phase, the second-order mode mixes the real and imaginary
parts of $a_1$ into the RF and DD signals, and the contamination varies not
only with the magnitude of $c_2$ but with its phase. In a high-power cavity
whose eigenmode changes thermally, the contamination itself therefore drifts
in time. Moreover, the effect is not small, since a mismatch of a few percent in power
corresponds to a $10$ to $20$\,\% contamination in amplitude.

These coefficients, like those of Eqs.~(\ref{eq:rfprime})--(\ref{eq:rho}),
keep $c_2$ to first order in the excitation. Since the next order shifts the fitted $c_2$ by
up to a few tens of percent (Sec.~\ref{sec:systematics}), here we give the
closed form to $O(c_2^2)$ that the fit uses.

Two effects enter at this order. The first is the fourth-order mode
content that the mismatch itself carries. In the eigenmode basis, the overlap of two Gaussian
beams~\cite{anderson1984,bayerhelms1984} gives a mismatched input beam that
contains only even modes,
\begin{equation*}
  u_{\mathrm{in}} = (1 - 2|c_2|^2)^{1/4} \sum_n
  \frac{\sqrt{(2n)!}}{2^n n!} \left(\sqrt{2}\, c_2\right)^n u_{2n},
\end{equation*}
whose $u_4$ coefficient gives
\begin{equation}
  c_4 = \frac{\sqrt{6}}{2}\, c_2^{\,2}.
  \label{eq:c4rel}
\end{equation}

The second is a uniform depletion of the carrier. The same expansion
carries a normalization $(1-2|c_2|^2)^{1/4} \simeq 1 - |c_2|^2/2$ common
to every mode, so the reflected field as a whole, carrier and sidebands
alike, is scaled by this factor. The bracket
$\langle E_c|M|E_\pm\rangle$ that produces every signal of this appendix
is bilinear in the field, so it picks up the square of the scaling,
$1 - |c_2|^2$. But $S_{\mathrm{DD,SUM}}$ is built from the same
reflected field and carries the identical factor, so the depletion
cancels exactly in every normalized observable $\hat{S}_{\mathrm{RF}}$,
$\hat{S}_{\mathrm{DD}}$ of this paper. It changes none of the
coefficients below.

What is left is the beat of the $u_4$ content against the excitation,
carried through the pairs $(u_0,u_3)$, $(u_0,u_5)$, and $(u_1,u_4)$ that the
quadrant mask admits at this order. Passing the excitation ladder
through the beat bookkeeping above extends Eqs.~(\ref{eq:rfcont}) and
(\ref{eq:ddcont}) to
\begin{align}
  \hat{S}_{\mathrm{RF}}^{(h)} &= \frac{Q_{01}}{\chi_a}
    \left\{ \left[ 1 + \frac{\tilde{c}_{2,h}^{\,r}}{\sqrt{2}}
      - \frac{\sqrt{6}}{12}\, \tilde{c}_{4,h}^{\,r} \right]
      \mathrm{Im}\!\left[ a_1\, e^{-i\eta_h} \right] \right. \nonumber \\
    &\hspace{4em} \left.
      {} - \left[ \frac{3}{\sqrt{2}}\, \tilde{c}_{2,h}^{\,i}
      - \frac{7\sqrt{6}}{12}\, \tilde{c}_{4,h}^{\,i} \right]
      \mathrm{Re}\!\left[ a_1\, e^{-i\eta_h} \right] \right\},
  \label{eq:rfcont2} \\
  \hat{S}_{\mathrm{DD}}^{(h)} &= -\, Q_{01}
    \left\{ \left[ 1 - \frac{3}{\sqrt{2}}\, \tilde{c}_{2,h}^{\,r}
      + \frac{7\sqrt{6}}{12}\, \tilde{c}_{4,h}^{\,r} \right]
      \mathrm{Re}\!\left[ a_1\, e^{-i\eta_h} \right] \right. \nonumber \\
    &\hspace{4em} \left.
      {} - \left[ \frac{1}{\sqrt{2}}\, \tilde{c}_{2,h}^{\,i}
      - \frac{\sqrt{6}}{12}\, \tilde{c}_{4,h}^{\,i} \right]
      \mathrm{Im}\!\left[ a_1\, e^{-i\eta_h} \right] \right\},
  \label{eq:ddcont2}
\end{align}
with $\tilde{c}_{4,h} \equiv c_4\, e^{-4i\eta_h}
= (\sqrt{6}/2)\, \tilde{c}_{2,h}^{\,2}$ by Eq.~(\ref{eq:c4rel}). The same
three pairs extend Eqs.~(\ref{eq:rfprime}) and (\ref{eq:ddprime}) to
\begin{align}
  \hat{S}_{\mathrm{RF}}^{(h)} &= -\, \frac{Q_{01}}{\chi_a}
    \left[ \sqrt{2}\, \tilde{c}_{2,h}^{\,i}
      - \frac{\sqrt{6}}{2}\, \tilde{c}_{4,h}^{\,i} \right]
    \mathrm{Re}\!\left[ a_1'\, e^{-i\eta_h} \right],
  \label{eq:rfprime2} \\
  \hat{S}_{\mathrm{DD}}^{(h)} &= -\, Q_{01}
    \left[ 2 - \sqrt{2}\, \tilde{c}_{2,h}^{\,r}
      + \frac{\sqrt{6}}{2}\, \tilde{c}_{4,h}^{\,r} \right]
    \mathrm{Re}\!\left[ a_1'\, e^{-i\eta_h} \right].
  \label{eq:ddprime2}
\end{align}
The three pairs enter Eq.~(\ref{eq:rfcont2}) with near cancellation in
one quadrature and reinforcement in the other, and enter
Eq.~(\ref{eq:rfprime2}) predominantly reinforcing, consistent with the
shift of Table~\ref{tab:bias}, which arises in the $a_1'$ responses.

Taking the same-sensor ratio [Eq.~(\ref{eq:rho})] with the full forms of
Eqs.~(\ref{eq:rfprime2}) and (\ref{eq:ddprime2}) gives
\begin{equation}
  \frac{\hat{S}_{\mathrm{RF}}^{(h)}}{\hat{S}_{\mathrm{DD}}^{(h)}}
  = \frac{1}{\chi_a}\,
    \frac{\sqrt{2}\, \tilde{c}_{2,h}^{\,i}
      - \frac{\sqrt{6}}{2}\, \tilde{c}_{4,h}^{\,i}}
         {2 - \sqrt{2}\, \tilde{c}_{2,h}^{\,r}
      + \frac{\sqrt{6}}{2}\, \tilde{c}_{4,h}^{\,r}},
  \label{eq:rho2}
\end{equation}
exact to $O(c_2^2)$ once Eq.~(\ref{eq:c4rel}) is substituted. Fitting this
ratio with the linear model of Eq.~(\ref{eq:rho}) instead would return a
$\tilde{c}_{2,\mathrm{meas}}$ that differs from the true $\tilde{c}_2$ (the
sensor index $h$ is dropped from here on).
Expanding Eq.~(\ref{eq:rho2}) to $O(c_2^2)$ and matching the linear
coefficient gives
\begin{equation}
  \tilde{c}_{2,\mathrm{meas}}^{\,i} = \tilde{c}_2^{\,i}
    \left( 1 - \sqrt{2}\, \tilde{c}_2^{\,r} \right),
  \label{eq:c2bias}
\end{equation}
independent of $\chi_a$ and of the carrier depletion above. The bias
itself, $-\sqrt{2}\, \tilde{c}_2^{\,r}\, \tilde{c}_2^{\,i}$, is quadratic
in $c_2$, so a run with twice the $|c_2|$ of another carries four times the
bias. Relative to the true $\tilde{c}_2^{\,i}$,
the bias is $-\sqrt{2}\, \tilde{c}_2^{\,r}$, linear in $|c_2|$ and
reaching $\sqrt{2}\, |c_2| \sim 8$ to $17\,\%$ for the mismatch of this
cavity. The fits of this paper use the second-order form throughout, so
this bias is absorbed rather than corrected for afterwards.
Its size in the fits of the main text, and what the next order would cost,
follow.

Truncating the fit at first order shifts the fitted $c_2$ as a whole
(Table~\ref{tab:bias}), $\mathrm{Im}[c_{20}]$ at $62$\,W by $20$\,\%, nearly
three times its statistical error, the other components by up to about
$10$\,\% and at most about two standard deviations, and the Gouy phases barely. The cause is the $u_4$ that the
mismatch itself carries, $c_4 = (\sqrt{6}/2)\,c_2^2$, which enters the $a_1'$
responses [Eqs.~(\ref{eq:rfprime})--(\ref{eq:rho})] at a relative size of
$(\sqrt{6}/2)\,|c_{20}| \approx 0.14$ for the $|c_{20}|^2 \approx 1.4$\,\%
here; the PIT plane, with a quarter of the mismatch power, shifts
proportionally less. The third order costs the same factor again,
$3$ to $4$\,\%, comparable to the $2.5$\,\% floor of
Sec.~\ref{sec:systematics}, and at a $1$\,\% power mismatch the
second and third orders fall to $10$\,\% and $1$\,\%, as the $2$\,W rows show.
A third-order correction is available by the same route, so the expansion
does not limit the measurement. Nor does it set the $\chi^2$ floor, which did
not move when the second order was added, the $a_1'$ ratios, two per plane,
being matched one for one by the two components of each run's $c_2$ and
leaving no residual.

\begin{table}[htbp]
\centering
\caption{Shift of the fitted $c_2$ when the second-order terms are added
to the model (second order minus first order). The significance in
parentheses is against the statistical error of the adopted fit.}
\label{tab:bias}
\begin{tabular}{llcc}
  \toprule
  Run & Plane & $\Delta\mathrm{Re}[c_2]$ & $\Delta\mathrm{Im}[c_2]$ \\
  \midrule
  $2$\,W & PIT & $-0.0024\;(0.4\sigma)$ & $+0.0006\;(0.1\sigma)$ \\
         & YAW & $-0.0043\;(0.7\sigma)$ & $+0.0038\;(1.0\sigma)$ \\
  $62$\,W run 1 & PIT & $-0.0022\;(0.8\sigma)$ & $+0.0019\;(0.5\sigma)$ \\
                & YAW & $+0.0003\;(0.1\sigma)$ & $+0.0134\;(2.8\sigma)$ \\
  $62$\,W run 2 & PIT & $-0.0030\;(2.2\sigma)$ & $+0.0017\;(0.5\sigma)$ \\
                & YAW & $-0.0014\;(0.4\sigma)$ & $+0.0138\;(2.8\sigma)$ \\
  \bottomrule
\end{tabular}
\end{table}

The other second-order products stay below the floor: the odd $u_3$ from
$c_2$ times the static first-order amplitude $a_s$ that the miscentering of the
beam on the sensor leaves, at
$\sqrt{3}\,a_s c_2 \approx 2\times10^{-3}$; the $u_2$ created by the drives,
whose static part biases $c_2$ by less than a percent; and the products with
the residual detuning, $\chi_0\, c_2 \sim 10^{-3}$. With the beam mode cleaned
upstream~\cite{rudiger1981,willke1998}, higher orders arise mainly through
the same mechanisms. None of this is stochastic, and none of it reaches the floor.

\section{Signal processing and fit model}
\label{app:proc}

\begin{figure}[htbp]
  \centering
  \includegraphics[width=\linewidth]{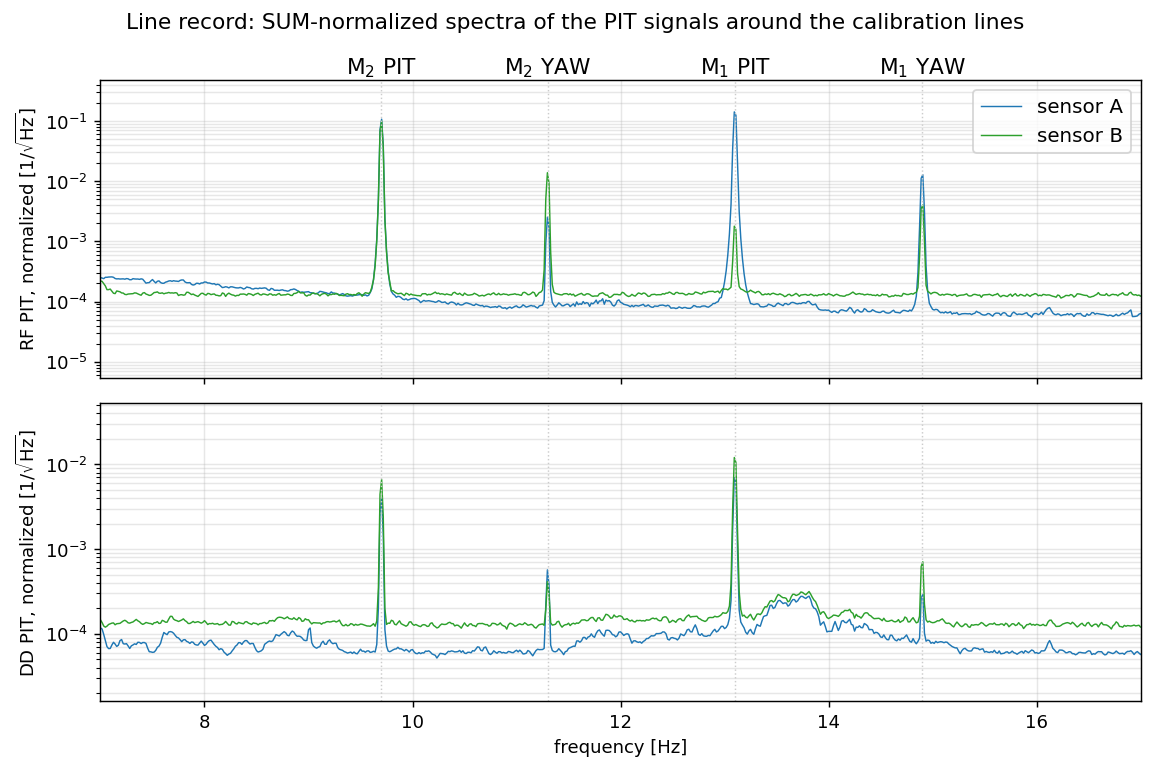}
  \caption{Line record: SUM-normalized spectra of the RF (top) and DD
    (bottom) PIT signals of both sensors, around the four calibration lines
    (dotted). The two PIT lines stand out in both chains, except the M$_1$
    PIT line in the RF signal of sensor B, which lies near a zero of that
    response since $\eta_B - \eta_1 \simeq -88^\circ$
    [Eq.~(\ref{eq:headratios})]; the YAW lines appear about an order of
    magnitude weaker through cross-coupling.}
  \label{fig:linespec}
\end{figure}

The processing is common to all records. Each segment's $I$/$Q$ pair is
first adjusted in demodulation phase. Since the reference-field line is a
pure length dither, each segment is rotated to remove the line from its
orthogonal quadrature, and the rotated $I$ signals are then combined into
PIT, YAW, and DIAG, following the mode indices of Sec.~\ref{sec:config}
($u_{20} \propto x^2$, $u_{02} \propto y^2$). The DD signals are formed by
numerical re-demodulation at the reference-field frequency, and everything is
normalized by the segment-summed DD signal [Eq.~(\ref{eq:ddsum})].
Figure~\ref{fig:linespec} shows the normalized spectra of the line record
around the four calibration lines. For scale, the sweep corresponds to a beam
motion of roughly $\pm 0.03$ of the beam radius on the sensor and the
calibration lines to $|a_1|$ of order $10^{-3}$, rough figures read from the
DD signals with the geometric projections ignored. None of them enters the
measurement, which is built from the normalized signals alone.

For the calibration lines, the fit uses the normalized line signals
themselves rather than their ratios, since a ratio acquires strongly
non-Gaussian errors whenever its denominator passes near zero, while the
signal errors are well behaved. Each line then brings in a scale degree of
freedom, the strength of its excitation, which the method never uses and
which is turned into a measurement instead. Each line is divided beforehand
by an independent estimate of the excitation it delivers, calibrated against
the transmitted power (Appendix~\ref{app:dither}), and the dither depth
likewise, the RF observables being multiplied by the $\chi_a$ from the same
route. The free parameters are then one scale $k$ per line and one
$k_\chi$ for the depth, each the ratio of the true quantity to its
estimate. Weak priors ($1 \pm 0.10$) hold the line scales near unity, and
none of the reported quantities depends on them, the physics living in the
ratios. The Gouy phase of each plane is carried as a phasor
$(\cos\eta, \sin\eta)$.

The model uses the full form of the response coefficients to second order in
$c_2$ (Appendix~\ref{app:coeff2}), with the residual detuning $\chi_0$ (defined
in Appendix~\ref{app:coeff}) taken as a known input, measured as
$-4.7 \times 10^{-3}$ in half-linewidth units. The $a_1$ calibration lines
were recorded only in the $2$\,W state, and the $62$\,W fits use that same
record for their geometry. In those fits the $c_2$ of the calibration-line
record floats as a separate parameter, the four picomotor-drive ratios are
absorbed exactly by the four $c_2$ components of the run itself, and the
geometry is determined by the lines alone. Keeping the calibration lines
running at $62$\,W would remove this complication, a single record then
determining everything at once.

\section{Implementation requirements and design guidelines}
\label{app:design}

\subsection{Hardware and excitation}

The method needs an EOM for the RF sidebands, an angular actuator before the
cavity, a length actuator to dither the cavity, an actuator that moves the beam
on the sensors in the reflected path, and two wavefront sensors. This is the
minimal configuration of a cavity with alignment control, with no element
added, so the method fits existing systems as they are.

The RF response to $a_1'$ is suppressed by $c_2$, so the drive should exceed the
WFS signal floor, the free beam motion or the sensor noise, by at least
a factor of $1/|c_2|$; with $|c_2|$ of order $0.1$ even at a $1$\,\% power
mismatch, ten times
the ordinary beam motion already gives ample signal.

The dither depth $\chi_a$, in half linewidths, is bounded from below by the DD
lines, which run below their RF counterparts by a factor of $\chi_a$ and must
stay above the segment noise floor. For a shot-noise-limited segment carrying
about one eighth of the detected power, that floor lies near $10^{-6}$ of the
half linewidth in one second for milliwatts on the diode, so a dither of
$10^{-4}$ to $10^{-3}$ of the half linewidth, picometers or less in length,
leaves orders of magnitude to spare. From above, $\chi_a \ll 1$ keeps the
modulation of the stored power at $2\omega_L$, of relative size $\chi_a^2/2$,
negligible. This demonstration used $\chi_a \simeq 0.09$ in the
calibration-line record, a $0.4$\,\% power modulation that left the operating
state undisturbed, and $\chi_a \simeq 0.004$ in the picomotor records. The frequency $\omega_L$ is largely free,
needing only a resolvable line that is harmless to the system.

The excitations need not run continuously. The $a_1$ measurement mainly carries
the actuator--sensor Gouy phases, which hardly drift, so after one
calibration-line measurement $a_1'$ alone gives a trustworthy $c_2$, and driving
$a_1'$ above the alignment-control bandwidth leaves the cavity essentially
unperturbed.

\subsection{Placement in Gouy phase}

The sensor placement departs from alignment-control practice. The alignment
optimum places the two sensors $90^\circ$ apart~\cite{morrison1994}, which for
$c_2$, rotating at twice the Gouy phase, becomes a degenerate $180^\circ$. With
the sensor separation $\Delta\eta_{AB} \equiv \eta_A - \eta_B$, the
conditioning goes as $|\sin\Delta\eta_{AB}|$ for the first order and
$|\sin 2\Delta\eta_{AB}|$ for the second, and the two are equal at
$\Delta\eta_{AB} = 60^\circ$, the optimum for alignment and mismatch together.
Actuators enter only at their Gouy phase $\eta_a$, not at $2\eta_a$, so any
two actuator planes not
$180^\circ$ apart close the system, $90^\circ$ being a natural choice.

The anchor, which fixes the common offset of the Gouy phases (Sec.~\ref{sec:apparatus}),
sits either near the cavity, with the distance to the waist converted to an
angle as in this demonstration, or in the far field, where the Gouy phase has
stopped accumulating; in both cases few lenses should lie between the anchor
and the cavity, since each adds astigmatism and model dependence. Finally, an $a_1'$
actuator placed before the split to the sensors drives both with a common
$a_1'$ and makes the inter-sensor ratio of Eq.~(\ref{eq:rfprimeratio})
available, and at the Gouy midpoint of the two sensors it gives a
readout of $c_2$ free of $\chi_a$; with per-sensor drives after the split, as
here, the same-sensor ratio of Eq.~(\ref{eq:rho}) is used.

\section{Calibration of the dither depths against the transmitted power}
\label{app:dither}

Every dither of this measurement, length and angular alike, is calibrated
against the transmitted power. On resonance the transmitted power responds
quadratically to either drive, so the line at twice a dither frequency returns
that dither's depth as a dimensionless ratio needing no calibrated detector,
no finesse, and no actuator model.

For the length, the detuning $\chi(t) = \chi_0 + \chi_a \sin \omega_L t$ in
half linewidths gives the transmitted power $P$, relative to its value
$P_{\mathrm{max}}$ on resonance, as
\begin{equation}
  \begin{split}
  \frac{P}{P_{\mathrm{max}}} = \frac{1}{1+\chi(t)^2}
  \simeq{}& 1 - \chi_0^2 - \frac{\chi_a^2}{2}
       - 2 \chi_0 \chi_a \sin \omega_L t \\
       & + \frac{\chi_a^2}{2} \cos 2\omega_L t .
  \end{split}
  \label{eq:trans2f}
\end{equation}
The line at twice the dither frequency does not involve the static detuning
$\chi_0$, so its ratio to the mean transmitted power is $\chi_a^2/2$ and returns
the depth directly. This is the same second-order effect that fixes the
upper bound on the dither depth in Appendix~\ref{app:design}, read here as a
measurement rather than as a limit. This $\chi_a$ is the depth that the
normalized signals carry (Sec.~\ref{sec:wfs}), so this route reads the fit's
quantity directly.

An angular dither is calibrated by the same response. A misalignment $a_1$
moves the fraction $|a_1|^2$ of the power into the first-order mode, which
the cavity does not transmit, so
\begin{equation}
  \frac{P}{P_{\mathrm{max}}} \simeq 1 - |a_1|^2,
  \label{eq:transang}
\end{equation}
and for a misalignment made of a static part and a dither
$a_d \sin \omega_a t$ the ratio of the line at
$2\omega_a$ to the mean transmitted power is $a_d^2/2$, returning the
excitation exactly as for the length. The excitation estimates that the
scales of Sec.~\ref{sec:fit} are measured against come from this
route, each read at twice its own dither frequency.

These estimates share one photodiode chain, so a common error in it moves
every scale of Table~\ref{tab:scales} together, and the scales sit $2$ to
$4$\,\% below unity. One such error lies in the whitening of the
transmitted-power chain, whose digital compensation uses the nominal design
response rather than that of the installed components. Toggling the whitening
during the length-dither record measures it directly: a calibrated depth reads
high by $(3.91 \pm 0.10)$\,\% with the whitening engaged, consistent with the
gain the components predict. Corrected for it, the dither-depth scale becomes
$1.00$, while the line scales, corrected also for the gap between the quadrant
segments, become $0.91$ to $0.96$, the difference discussed in
Sec.~\ref{sec:actcal}.

\section{Beam centering with the DD signal}
\label{app:centering}

In the alignment error signals, the contamination of Sec.~\ref{sec:fit} and
Appendix~\ref{app:coeff2} causes two kinds of harm. The change that $c_2$ makes
to the coefficients through which the sensors read $a_1$
[Eqs.~(\ref{eq:rfcont}) and (\ref{eq:ddcont})] is a calibration
change: the sensing matrix rotates and rescales and drifts with the thermal
state, explaining part of the drift so far handled empirically, but a
recalibration from the measured $c_2$ and Gouy phases absorbs it. The coupling
of Eq.~(\ref{eq:rfprime}) is different: through $c_2$ the RF alignment signal
acquires a sensitivity to beam position, and miscentering leaks into the
alignment loop as a spurious error signal. The first prescription is therefore
independent of $c_2$: center the beam, and hold it centered, with the DD
signal as the error signal.

The SUM-normalized DD difference [Eq.~(\ref{eq:ddprime})] is a calibrated
beam-centering signal: it reads the spot position on the quadrant
photodetector in units of the
beam radius, with the coefficient $-2Q_{01}$ fixed by geometry alone. Every
factor that depends on the state of the cavity (the audio-sideband
amplitude, the reflectivities, the optical power) is common to the numerator
and the denominator and cancels in the ratio, so the calibration holds
independently of the coupling and resonance state. The DC centroid offers no
such signal in the resonant state: near critical coupling the reflected
carrier $u_0$ is suppressed, and the DC centroid is heavily contaminated by
the second-order content of the reflected field. The DD centering signal is
immune to this contamination at leading order, because its carrier-side beat
partners are the audio sidebands, which ride on $u_0$ alone.

\section{Cavity parameters}
\label{app:cavity}

\begin{table}[H]
\centering
\caption{Parameters of the cavity (unit SN008) used in the demonstration.
The mirror transmissions, per-mirror loss, and input/output asymmetry are the
bench-measured power-budget values of Ref.~\cite{nakano2025jac}; the finesse
and the free spectral range were measured in the same campaign.}
\label{tab:cavity}
\begin{tabular}{ll}
  \toprule
  Geometry & bow-tie ring \\
  Straight-section length & $0.5$\,m \\
  Width & $101.6$\,mm \\
  \midrule
  \multicolumn{2}{l}{\emph{Input and output couplers} ($2$)} \\
  \quad Radius of curvature & flat \\
  \quad Transmission & $2.4930$\,\% \\
  \quad Differential asymmetry & $-0.002$ \\
  \midrule
  \multicolumn{2}{l}{\emph{High reflectors} ($2$)} \\
  \quad Radius of curvature & $3$\,m \\
  \quad Transmission & $75$\,ppm, \; $74$\,ppm \\
  \midrule
  Loss per mirror & $17$\,ppm \\
  Free spectral range & $148.38$\,MHz \\
  Finesse & $124.8$ \\
  Waist radius, horizontal & $546.3$\,\textmu m \\
  Waist radius, vertical & $549.0$\,\textmu m \\
  \bottomrule
\end{tabular}
\end{table}

\bibliography{references}

\end{document}